\documentstyle[12pt]{article}
\begin{document}
\hoffset=-1truecm
\voffset=-2truecm
\baselineskip=18pt plus 2pt minus 2pt
\hbadness=500
\tolerance=5000
\title{\bf On the theory of magnetic impurities in integrable
correlated electron chains}
\author{A.A. Zvyagin \\
B.I. Verkin Institute for Low Temperature Physics and Engineering, \\
Ukrainian National Academy of Sciences, 47 Lenin Avenue, \\
Kharkov, 61164, Ukraine}
\date{April 15, 2002}
\maketitle

In a recent preprint \cite{prep} Ge {\em et al.} reply to our recent
response \cite{an} to some concerns raised in \cite{Ge}. Several obvious
remarks are necessary to clarify the situation, because the authors of
\cite{prep} certainly misunderstand (or misinterpret) some of our 
statements. 

1. In \cite{Ge} and in \cite{NP} the authors imply that integrable
magnetic impurities cannot exist in closed $t\!-\!J$ and Hubbard
chains (cf. P.~795 of \cite{NP} and P.~8544 of \cite{Ge}). It turns
out that in \cite{prep} they admit that integrable magnetic impurities
{\em can} exist in exactly solvable closed correlated electron chains. 

2. The approach (i) in our answer \cite{an} has been named the ``quantum 
inverse scattering method'' (also known as the algebraic Bethe ansatz
cf. \cite{KBI}) long before our work on impurities in correlated 
electron systems, see, e.g., \cite{rev}. This standard definition has
been used in our reply \cite{an}. The approach (ii) in our reply 
\cite{an} is known as the ``graded quantum inverse scattering method'' 
\cite{EK}. Does the strong emphasis in \cite{prep} on the difference 
between the co-ordinate Bethe ansatz and the quantum inverse scattering 
method imply that the authors of \cite{prep} believe that those two 
methods could yield {\em different} answers? We are not aware of any 
such contradictions between the two methods. 

3. In \cite{an} we pointed out that the impurity matrix changes the 
commutation relations in the spin sector (see below). This is 
absolutely correct, keeping in mind that two parameters, $\theta$ and
$S$, which distinguish the impurity site from other sites of the chain
are nonzero (note that the impurity scattering matrix used in our
papers \cite{SchZv97,Zv97,Zv99} mixes the states with $S$ and $S
+{1\over2}$; this hybridization is sometimes misunderstood).  

4. The magnetic impurities we studied in our papers have an essentially
different structure than those of \cite{oth}; hence, it is no wonder 
that the solutions do not coincide with ours. Ref.~\cite{Go} considers
the supersymmetric $t\!-\!J$ model with a different grading than the
one considered by us and does not consider magnetic impurities. Since
this represents a very different situation, it does not contradict our 
results. Actually, the special case of the impurity of \cite{F}
coincides with our results \cite{Zv99}.

5. In the approach (ii) for the supersymmetric $t\!-\!J$ model the
operators ${\hat A}_{12}$, ${\hat A}_{13}$, ${\hat A}_{21}$ and 
${\hat A}_{23}$ acting on the vacuum state {\em do} indeed yield zero
(cf. Eq.~(3.27) of \cite{EK}) in the FFB grading, contrary to the 
statements in \cite{prep}. One can see that the results of \cite{Zv99} 
for the special case of $\theta =0$ and spin, equal to the ones of the
host, coincide with those of \cite{EK} (cf. Eq.~(3.50) of \cite{EK}
and (A1) of \cite{Zv99}). The operators ${\hat A}_{12}$ and ${\hat
A}_{21}$ in \cite{prep} do not contain any characteristics of the impurity
(i.e., $\theta$ and $S$), and are then equivalent to those studied in 
\cite{EK}. This way, the argumentation of \cite{prep} can be applied
to paper \cite{EK}, and the criticism presented in \cite{prep}
actually concerns \cite{EK} rather than \cite{Zv99} (in which we
essentially used the method developed in \cite{EK}). However, the
criticism presented in \cite{prep} is incorrect, because the authors
do not take into account the fact that the eigenvalue of the transfer
matrix is determined up to some multiplier \cite{KBI,EK,F}. Moreover, the 
important commutation relations for the spin sector are those 
between the ${\hat A}_{ij}$ ($i,j={1,2}$) and ${\hat A}_{31}$ and 
${\hat A}_{32}$ (or $C_{1,2}$, cf. \cite{EK}), and those between the 
latter two operators, which indeed are used as ``creation operators'' 
in \cite{Zv99}. Namely, the changes in these commutation relations, 
but not in those between ${\hat A}_{12}$, ${\hat A}_{13}$, 
${\hat A}_{21}$ and ${\hat A}_{23}$ (which are mentioned in \cite{prep}), 
determine the changes in the spin sector of Bethe ansatz equations due
to the magnetic impurity. 

6. The change of the ``class of the representation'' ($l$) implies the
change of the symmetry in the considered model (we did not discuss the
symmetry of the Lax operator in \cite{an}, however, it turns out that
the symmetries of our impurity $L$-operators and those of the host are
the same, unlike the case of Refs.~\cite{oth}). Hence, our statement
in response to \cite{Ge} is correct.  

7. Point (5) of our answer to \cite{Ge} pertains to approach (i), 
but not to approach (ii). However, in \cite{Zv99} the approach (ii)
was used. It is, naturally, correct \cite{prep} that in the FFB
grading of the approach (ii) one cannot use ${\hat A}_{21}$ as a 
``raising operator''. But the authors of Ref.~\cite{prep}
misunderstand our statements \cite{an} and incorrectly mix the two 
approaches. 

8. Obviously, the statements of our answer \cite{an} (and the results 
of our previous papers) do not contradict \cite{FS}. 

9. The claim in \cite{prep} that Ge {\em et al.} studied a {\em spin} 
impurity, without additional charge degrees of freedom, contradicts
the fact that to according their Bethe ansatz equations, e.g., derived in 
\cite{NP}, the valence (the occupation number at the impurity site) 
{\em varies} with external parameters (such as the chemical potential, 
a global (non-local) magnetic field), cf. \cite{Zv99}. This is 
impossible if one studies a pure magnetic impurity, which has only 
spin degrees of freedom (in this case the valence should be one 
and not vary with the external parameters, even for $q=1$).    

Discussions with P.~Schlottmann are acknowledged.

\vfill
\eject

\end{document}